# Emergent superconductivity and pair density wave at antiphase boundaries of charge density wave order in kagome metals

Xianghe Han[1,2]†, Hui Chen[1,2]*†, Hengxin Tan[3], Zhongyi Cao[1,2], Zihao Huang[1,2], Yuhan Ye[1,2], Zhen Zhao[1,2], Chengmin Shen[1,2], Haitao Yang[1,2], Binghai Yan[3], Ziqiang Wang[4]*, Hong-Jun Gao[1,2]*

**Affiliations:**

[1]Beijing National Center for Condensed Matter Physics and Institute of Physics, Chinese Academy of Sciences, Beijing 100190, PR China

[2]School of Physical Sciences, University of Chinese Academy of Sciences, Beijing 100190, PR China

[3]Department of Condensed Matter Physics, Weizmann Institute of Science, Rehovot, Israel

[4]Department of Physics, Boston College, Chestnut Hill, MA 02467, USA

\* Corresponding author. Email: hjgao@iphy.ac.cn, wangzi@bc.edu, hchenn04@iphy.ac.cn

† These authors contributed equally to this work

**Abstract:** Central to the layered kagome lattice superconductors $A$V$_3$Sb$_5$ ($A$ = K, Cs, Rb) is a cascade of novel quantum states triggered by an unconventional charge density wave (CDW) order. The three-dimensional (3D) order involves a 2×2×2 phase coherent stacking of 2×2 charge density modulations in the kagome plane at low temperatures, exhibiting a CDW energy gap and evidence for time-reversal symmetry breaking. Here we report the discovery of emergent superconductivity and primary pair density wave (PDW) at the antiphase boundaries and stacking faults of bulk CDW order. We find that the $\pi$-phase shift dislocations can naturally appear on the surface as the Cs atoms form 2×2 superstructures that are out of phase with the bulk CDW. An incipient narrow band of surface states inside bulk CDW gap emerge close to the Fermi level where a particle-hole symmetric energy gap develops. We demonstrate that the energy gap originates from a novel quasi-2D kagome superconducting state ($T_c$ ~ 5.4 K) intertwined with bulk CDW order, exhibiting an unprecedented vortex core spectrum and spatial modulations of the superconducting gap consistent with a 4×4 PDW. Intriguingly, the 2D kagome superconductivity is shown to be tunable on and off by atomically manipulating the Cs atoms on the surface. Our findings provide fresh new insights for understanding the interplay between the unconventional CDW and superconductivity in kagome metals and a pathway for atomic manipulation and topological defects engineering of quantum many-body states in correlated materials.

The vanadium-based kagome metals $A$V$_3$Sb$_5$ ($A$ = K, Cs, Rb) have attracted enormous interests in the field of quantum materials(*1-12*). The bulk single-crystals have a layered structure containing two-dimensional (2D) kagome lattice planes formed by V atoms (Fig. 1A)(*1, 2*). They are unique because $A$V$_3$Sb$_5$ are nonmagnetic and have a superconducting (SC) ground state(*2, 13*), in contrast to the insulting(*14*) and itinerant magnetic(*15, 16*) transition-metal kagome materials. At the center of the phenomenology is a cascade of symmetry breaking correlated and topological quantum states triggered by an unconventional charge-density wave (CDW) order(*4, 17-21*). In additional to the broken lattice translation symmetry, rotation symmetry is broken in an intricate manner(*22*), exhibiting a rich set of electronic liquid crystal states such as smectic order(*5, 23, 24*), stripes(*9, 25*), and electronic nematicity(*7, 26, 27*). There are also evidence for spontaneously time-reversal symmetry breaking(*4, 11, 12, 28*), which is highly unexpected and nontrivial in the absence of spin magnetism and under intensive debate(*23, 29, 30*). These many-body states transition to a novel SC state at low-temperatures that exhibits primary pair-density wave (PDW) order(*6*), pseudogap behavior(*6*), and charge-6$e$ superconductivity in thin-film ring structures(*31*). Understanding the physical origin and the broader implication of these novel quantum phenomena is currently the focus of intensive research(*32, 33*). A fundamental question of overarching importance is the interplay between the unconventional CDW and superconductivity, and whether these or new quantum phases can be achieved in the 2D limit of the kagome lattice(*34*). This is challenging to address directly since a metal-insulator transition has been observed when the thickness of exfoliated films is less than five layers(*35*).

The CDW state in AV$_3$Sb$_5$ is three-dimensional (3D) with strong inter layer correlations. It can be thought as a phase coherent stacking along the $c$-axis of $2a_0 \times 2a_0$ (2×2 in short) charge density modulations in the kagome plane. The stabilized 2×2×2 order at low temperatures creates a bulk CDW energy gap in the electronic density of states as shown schematically (Fig. 1B) for CsV$_3$Sb$_5$. Performing low-temperature scanning tunneling microscopy/spectroscopy (STM/S) on naturally cleaved surface of CsV$_3$Sb$_5$, we discovered large regions over which the Cs atoms form 2×2 superstructures commensurate with the charge density modulations in the V kagome plane. Remarkably, the 2×2 Cs order can be out of phase with the bulk CDW, causing a π-phase shift dislocation stacking fault on the surface where a narrow band of surface states appears inside the bulk CDW gap, giving rise to a prominent spectral peak in the tunneling density of states close to the Fermi level (Fig. 1B, right panel). Intriguingly, a particle-hole symmetric energy gap develops at low temperatures well above the bulk SC transition (Fig.1B, right panel). By a combination of systematic studies of the temperature dependence, magnetic field evolution, and magnetic vortex imaging, we identify the energy gap with an emergent quasi-2D SC state from the band of bound states localized near the surface with stacking dislocations. This novel 2D superconductivity is distinct from the bulk SC state, and has significantly higher critical temperature ($T_c \sim$ 5.4 K), out-of-plane critical magnetic field ($H_{c2} \sim$ 8T), and two-gap to $T_c$ ratio (~7.3) than those associated with the bulk superconductivity in CsV$_3$Sb$_5$. Moreover, the new SC state is highly unusual, exhibiting previously unseen spectrum for vortex core excitations and 4×4 spatial modulations of the SC coherence peak and the SC gap size, consistent with a 3Q primary PDW.

## 2×2 order of Cs atoms on surface

The CsV$_3$Sb$_5$ single crystals are cleaved at low temperature (details see Methods in Supplementary Materials). Different from the well-known Cs reconstructions on cleaved surfaces reported previously (Fig. S6), we observe a new type of ordered Cs structure in some surface regions (Fig. 1C and Fig. S8). The STM topographic image and its Fourier transform (FT) reveal a hexagonal lattice with the periodicity about 1.1 nm, twice of the crystalline lattice constant $a_0$, demonstrating the 2×2 ordered superlattice structure (Fig. 1C). To further reveal the atomic configuration of Cs atoms, we intentionally push some of them away by STM tip to expose a small-area of the Sb2 surface below(6) (Fig. 1D). We identify that each Cs atom in the 2×2 superstructure locates right above the center of the Sb2 honeycomb lattice, which is directly above the Sb1 atom at the center of the hexagons in the V kagome lattice (Fig. 1A). The observations of 2×2 Cs superstructures are reproducible on all samples cleaved at low temperature. The largest area of 2×2 Cs ordered surface is about 150 nm × 150 nm (details see Methods in Supplementary Materials).

## Incipient surface band inside bulk CDW gap

We next study the electronic properties on the 2×2 ordered Cs surfaces by collecting spatially-averaged differential conductance (d$I$/d$V$) spectra (Fig.1E). In sharp contrast to other types of surfaces such as √3×√3 ordered or the Sb2 surface where the spectra show a density of states suppression over a broad soft CDW energy gap, a series narrow bands of bound states emerge inside the bulk CDW gap on the 2×2 ordered Cs surfaces, with the sharpest spectral peak located at about 5.6 mV ($P_{2D}$) above the $E_F$ (Fig. 1E). The full width at half maximum of the peak is about 2 mV and its intensity is nearly five times larger than the background tunneling conductance from the bulk bands (Fig. 1E, and F). These incipient band of near mid-gap states are unexpected, especially considering the fact that the 2×2 ordered Cs atoms have the same planar periodicity as the bulk 2×2×2 CDW order, indicating their intricate interplay to which we will return below.

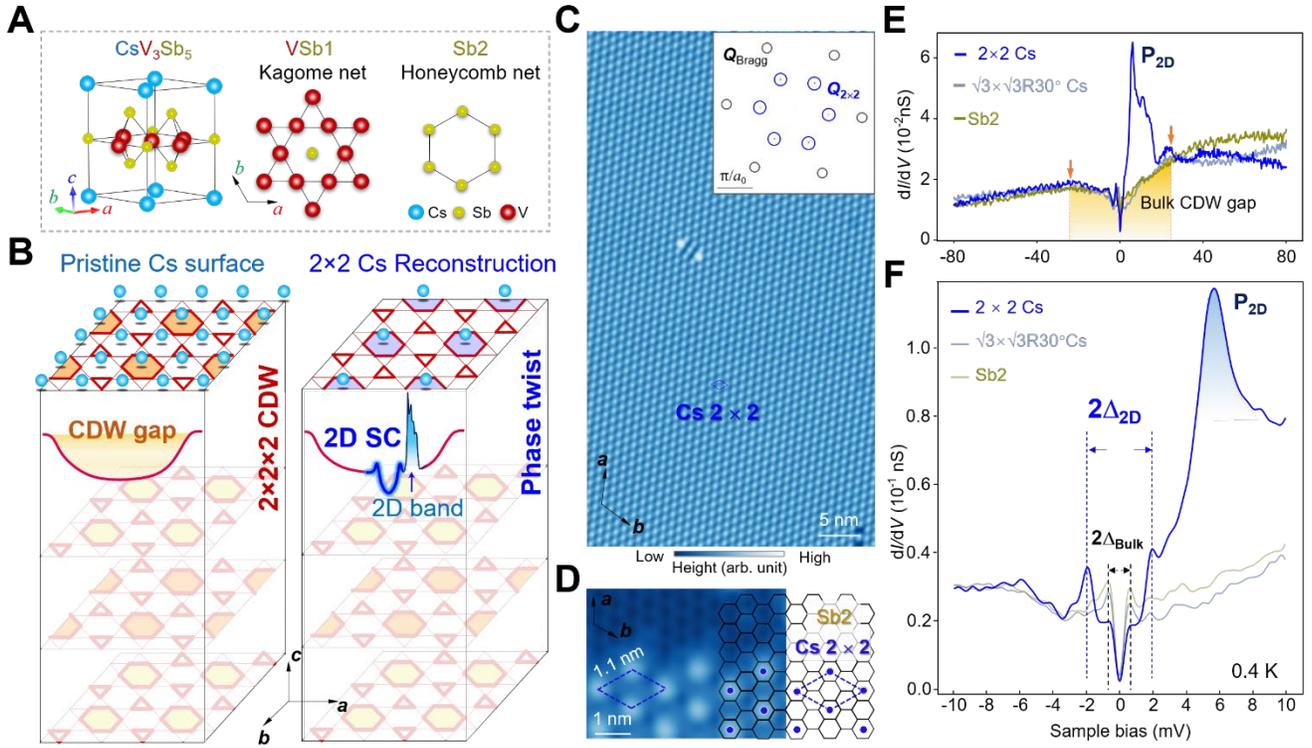

**Fig. 1. Emergent low-energy gap and incipient surface band inside the bulk CDW gap on the 2×2 Cs ordered surface of CsV$_3$Sb$_5$.** (**A**) The atomic structure of CsV$_3$Sb$_5$, exhibiting a layered structure of V-Sb sheets intercalated by Cs. In the VSb1 layer, the kagome net of vanadium is interwoven with a simple hexagonal net formed by the Sb1 atoms. The Cs atoms sit right above the Sb1 atoms, forming hexagonal lattice. The Sb2 layer forms honeycomb lattice. (**B**) Schematic of the 2×2 ordered Cs surface of CsV$_3$Sb$_5$, showing the emergence of quasi-2D SC gap and a narrow band of bound states when the 2×2 Cs order is out of phase with bulk CDW. (**C**) STM image of as-cleaved CsV$_3$Sb$_5$ at low temperature showing a large area of 2×2 ordered superlattice structure ($V_{set}$= 900 mV, $I_t$=50 pA). Inset: the Fourier transform of the STM image, showing the **Q**$_{2\times2}$ (blue circles) and **Q**$_{Bragg}$ (black circles). (**D**) STM image (left) and schematic (right) of the 2×2 Cs (blue dashed rhombus) and Sb surface regions, showing the atomic configuration of Cs atoms (blue spots) on the honeycomb lattice of Sb2 layer (black hexagons). $V_{set}$= 40 mV, $I_t$= 1 nA. (**E, F**) The spatially-averaged d$I$/d$V$ spectra obtained on different surfaces of CsV$_3$Sb$_5$ at a temperature of 0.4 K in the voltage range of [-80 mV, 80 mV] (**E**) and [-10 mV, 10 mV] (**F**), respectively, showing the emergence of a series narrow bands of bound sates with the sharpest spectral peak around 5.6 mV ($P_{2D}$) inside the bulk CDW gap (highlighted by the orange shade) and a particle-hole symmetric gap of 1.9 meV ($\Delta_{2D}$) beyond the bulk SC gap ($\Delta_{bulk}$) at 2×2 Cs ordered surface. $V_{set}$= 100 mV, $I_t$= 3 nA, $V_{mod}$= 0.1 mV.

## Emergent low-energy gap at the Fermi level

The tunneling conductance at low energies shows the significantly enhanced density of states near the $E_F$ elevated by the tail end of the in-gap states peaked at $P_{2D}$ exclusively on the novel 2×2 ordered Cs surfaces (Fig. 1E). Intriguingly, a new pair of conductance peaks develop at around ±1.9 mV, delineating an emergent particle-hole symmetric gap labeled as $\Delta_{2D}$ (Fig. 1E). From hundreds of d$I$/d$V$ spectra collected over different 2×2 Cs ordered regions of six CsV$_3$Sb$_5$ samples, we obtain the average gap size of $\Delta_{2D}$ ~1.70±0.17 meV (Fig. S1A). This is in sharp contrast to other types of surfaces, where only the bulk SC gap $\Delta_{Bulk}$ ~ 0.5 meV was observed below the bulk $T_{c,bulk}$ ~ 2.5 K that is indeed present as a distinct in-gap feature inside $\Delta_{2D}$ on the newly-discovered 2×2 Cs ordered surfaces (Fig. 1E). The d$I$/d$V$ maps covering the boundary between 2×2 Cs ordered region and other types of Cs reconstructions show that the dominant intensity at 2 mV and 5 mV are localized in the 2×2 Cs ordered region, whereas no difference is observed across the boundary at zero energy (Fig. S1B-E). These observations further demonstrate that the $\Delta_{2D}$ and $P_{2D}$ emerge from and are characteristic of the quasi-2D surface states on the 2×2 Cs ordered surfaces.

## A new quasi-two-dimensional SC state

To investigate the origin of the emergent energy gap, we study the temperature and magnetic field dependence of $\Delta_{2D}$. Applying external magnetic fields perpendicular to the sample surfaces, we observe that the bulk SC gap $\Delta_{bulk}$ is fully suppressed above the upper critical field $H_{c2,bulk}$ ~ 0.8 T(*36*) (Fig. 2A), while $\Delta_{2D}$ survives and becomes undetectable only at a much larger critical field of $H_{c2,2D}$ ~ 8.0 T. In addition, the suppression of $\Delta_{2D}$ is independent with the perpendicular field direction (Fig. 2B), excluding the possibility of magnetic surface states in 2×2 Cs ordered regions for the nonmagnetic CsV$_3$Sb$_5$ crystal. Next, increasing temperature in the absence of an external magnetic field, the bulk superconducting gap $\Delta_{bulk}$ closes and is undetectable above a critical temperature of $T_{c,bulk}$ ~ 2.3 K(*6*), whereas the quasi-2D energy gap $\Delta_{2D}$ persists up to $T_{c,2D}$ ~ 5.4 K (Fig. 2C, and D). The peak-to-peak gap value of $\Delta_{2D}$ (Fig. 2E) exhibits a similar temperature dependence as the bulk superconducting gap $\Delta_{bulk}$(*37*). Thus, the field and temperature dependent measurements suggest that $\Delta_{2D}$ most likely originates from an intrinsic quasi-2D superconductivity rather than a proximity effect induced superconductivity. The value of the measured gap-to-$T_c$ ratio $2\Delta_{2D}/k_BT_c$ ~ 7.3 is significantly larger than that of bulk superconductivity (~5.2)(*6*).

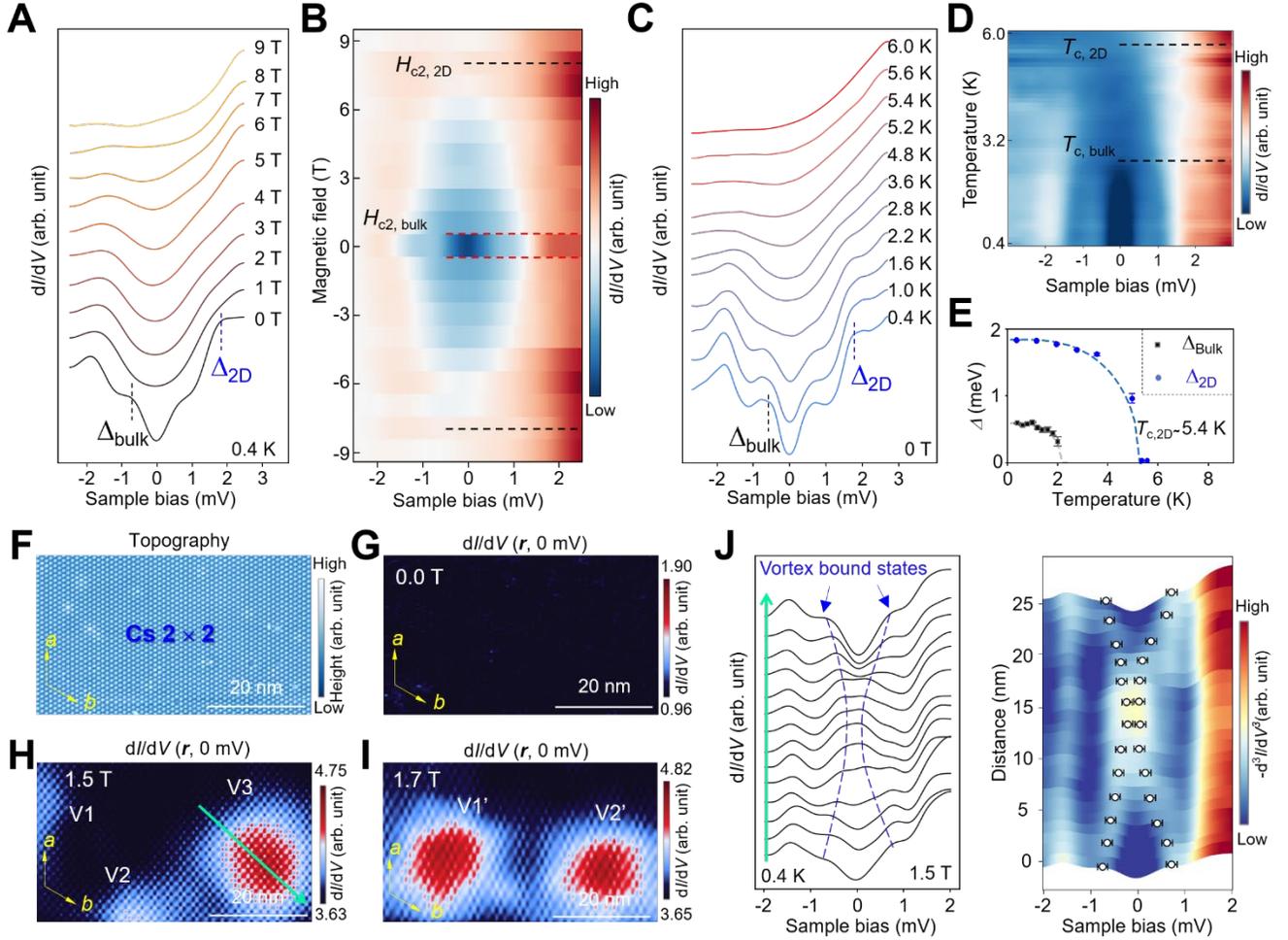

**Fig. 2. Superconductivity origin of $\Delta_{2D}$ at the 2×2 Cs ordered surface of CsV$_3$Sb$_5$.** (**A**) The out-of-plane magnetic field dependent d$I$/d$V$ spectra, showing the suppression of $\Delta_{bulk}$ and $\Delta_{2D}$. (**B**) Color map of d$I$/d$V$ spectra from -9 T to 9 T, showing the magnetic field induced suppression of $\Delta_{2D}$, independent with the field direction. The $H_{c2}$ of $\Delta_{2D}$ ($H_{c2,2D}$) is about 8 T (black dash line), which is much higher than the $H_{c2,bulk}$ of $\Delta_{bulk}$ (0.8 T, red dashed line). (**C, D**) Temperature dependent d$I$/d$V$ spectra (**C**) and corresponding color map (**D**), showing that the critical temperature of $\Delta_{2D}$ is significantly higher than the bulk SC gap. (**E**) Evolution of $\Delta_{bulk}$ and $\Delta_{2D}$ with temperature, showing the bulk superconductivity with $T_{c,bulk}$ = ~2.3 K and $T_{c,2D}$ = ~5.4 K. The blue dashed curve is the BCS simulation of SC gap size evolution. (**F-I**) STM image of a 2×2 Cs ordered surface (**F**, $V_{set}$= 900 mV, $I_t$ = 50 pA) and corresponding zero-energy d$I$/d$V$ maps under out-of-plane magnetic field of 0.0 T (**G**), 1.5 T (**H**) and 1.7 T (**I**), respectively, showing field-induced Abrikosov vortex lattice from $\Delta_{2D}$. The vortices in (**H**) and (**I**) are labelled as V$_1$-V$_3$ and V$_1$'-V$_2$' respectively. (**J**) The waterfall plot of d$I$/d$V$ spectra (left) and corresponding intensity plot of second derivative -d$^3I$/d$V^3$ spectra (right) along the green arrow in (**H**), showing the bound states inside $\Delta_{2D}$ across the vortex core. For all d$I$/d$V$ data in this figure, $V_{set}$= 50 mV, $I_t$ = 2 nA, $V_{mod}$ = 0.1 mV. The spectra in (**A, C, J**) are vertically shifted for clarity. The error bars in (**E**) represent the standard errors obtained by Gaussian fits shown in (**D**). The error bars in (**J**) represent the energy resolution in STS measurements.

**Novel vortex core states of the quasi-2D SC state**

The superconductivity origin of $\Delta_{2D}$ is further confirmed by the observation of Abrikosov vortices (Fig. 2F-J) in a perpendicular magnetic field $B_z$. At 0.4 K and $B_z$ = 0.0 T, the zero-bias conductance (ZBC) map is essentially uniform (Fig. 2G and Fig. S9). When $B_z$ is well above the bulk $H_{c2,bulk}$ ~ 0.8 T, $\Delta_{bulk}$ is fully suppressed and the vortex lattice of the bulk SC state is undetectable(*38*) (Fig. S10). However, several bound state cores with high spectral intensity labelled as V1, V2 and V3 (Fig. 2H) are observed at 1.5 T in the ZBC map of the 2×2 Cs ordered region where $\Delta_{2D}$ is present. The comparison between the ZBC map at 0 T and 1.5 T of the same surface region rules out the local impurity-induced bound states as the origin. When $B_z$ is increased to 1.7 T, the core positions change with the field (marked by V1' and V2' in Fig. 2I). The distance between two neighboring vortex cores decreases from 40 nm at 1.5 T to about 21 nm at 3.5 T (Fig. S11), demonstrating the formation of the Abrikosov vortex lattice(*38*). In addition, the d$I$/d$V$ linecut across the core shows that the emergence of vortex bound states inside $\Delta_{2D}$ (Fig. 2J). These observations demonstrate the existence of field-induced vortex lattice arising from the quasi-2D superconductivity associated with the energy gap $\Delta_{2D}$. A rough estimate of the vortex core size gives a coherence length of 6-8 nm, which is about three times smaller than the coherence length of the bulk SC state subtracted from bulk vortex core size ~ 26 nm (Fig. S12)(*18*). Since the critical field scales with the inverse of the coherence length squared, we arrive at a critical field that is about 10 times larger than that of the bulk SC states, which is indeed consistent with our observation of $H_{c2,2D}$ ~ 8 T discussed above. We note that the vortex bound states exhibit a plateau-like density of states around zero-energy (Fig. 2J), which is distinct from the conventional Caroli-de Gennes-Matricon states observed in bulk SC state of CsV$_3$Sb$_5$(*38*) (Fig. S12).

**Pair density wave modulations of $\Delta_{2D}$**

The emergent quasi-2D superconductivity may harbor intertwined quantum states invisible in the 3D compounds. We have thus studied the spatial distribution of the SC properties and observed the emergence of a novel 4×4 spatial modulation, which are twice the periodicity of the 2×2 Cs order, directly visible in the d$I$/d$V$ maps (white dotted circle in Fig. 3A). They give rise to three prominent peaks located at half the length of $\mathbf{Q}_{2×2}$ in the FT pattern (green circles in Fig. 3B). The new short-range ordered 4×4 modulations emerge only at low energies (e.g. -1.6 meV, Fig. 3A) while become undetectable at relatively higher energies (e.g. 6.4 meV, Fig. 3C, D). The FT intensity along one lattice direction (marked by the red dash line in Fig. 3B) as a function of sample bias shows that the modulation wave vector, i.e. $\mathbf{Q}_{4×4}$ does not depend on the bias energy and is nondispersive (Fig. 3E). In addition, $\mathbf{Q}_{4×4}$ is observed over the entire energy range of the density of states buildup due to the quasi-2D SC coherence (right panel of Fig. 3E), which is different from the modulations caused by the quasiparticle interference due to impurity scattering that should be limited to the very narrow energy window of the coherence peaks(*39*). These observations are reproducible in the different 2×2 Cs ordered regions of various samples (Fig. S13) and support that the 4×4 ordered modulations, the first of its kind observed in kagome superconductors, is correlated to and a part of the emergent quasi-2D superconductivity.

To directly test this picture, we further study the spatial variation of the quasi-2D SC gap size $\Delta_{2D}$ by collecting the d$I$/d$V$ spectra along a line-cut (green arrow in Fig. 3A), which show sizable variations in the distance between the pair of coherence peaks associated with $\Delta_{2D}$ (Fig. 3F). The local gap sizes $\Delta_{2D}$ can be more precisely extracted from the peak locations in the -d$^3I$/d$V^3$ curves as marked by the green crosses (Fig. 3G), which exhibit systematic modulations across the line-cut. The spatial modulation of $\Delta_{2D}$ have an amplitude of the order of 5.5% of the average gap. Intriguingly, the spatial modulations can be well described by two distinct periods of 2$a_0$ and 4$a_0$ respectively (green dashed curve in Fig. 3G), corresponding to an intra-unit-cell 2×2 pair density modulation (PDM)(*40, 41*) and a primary 4×4 PDW. We thus attribute both the PDM and the PDW formation to the emergent quasi-2D superconductivity.

It is instructive to compare to the 3**Q** PDW with 4/3$a_0$ × 4/3$a_0$ spatial modulations previously observed(*6*) in CsV$_3$Sb$_5$. Both **Q**$_{4×4}$ and **Q**$_{4/3×4/3}$ have been predicted theoretically for possible emergent density waves based on the near nesting of the Chern Fermi pockets in the 2×2 CDW state on the kagome lattice(*42*). There are two main differences between the **Q**$_{4×4}$ and the **Q**$_{4/3×4/3}$ PDW formation. First, the **Q**$_{4×4}$ PDW is observed in the 2×2 Cs ordered regions and is born out of the quasi-2D SC state on the surface. The **Q**$_{4/3×4/3}$ PDW, on the other hand, is a part of the bulk SC state observed on Sb surfaces after pushing the Cs atoms away using the STM tip. Second, the **Q**$_{4×4}$ PDW exists outside the bulk SC gap and hovers over the entire energy range of the spectral buildup caused by the quasi-2D SC coherence (Fig. 3E), whereas the **Q**$_{4/3×4/3}$ PDW exists over a large energy region defined by a pseudogap much larger than the bulk SC gap(*6*). While a microscopic mechanism behind the PDW formation is lacking, the difference may originate from the nesting of the incipient narrow bands of the 2×2 Cs ordered surface that directly give rise to the **Q**$_{4×4}$ vectors without relying on the CDW induced band folding(*42*). In addition, we stress that the new 3**Q** PDW with 4×4 modulations is unrelated to the unidirectional (1**Q**) 4$a_0$ charge stripe order observed on the Sb terminated surfaces over a large energy range(*5, 43*).

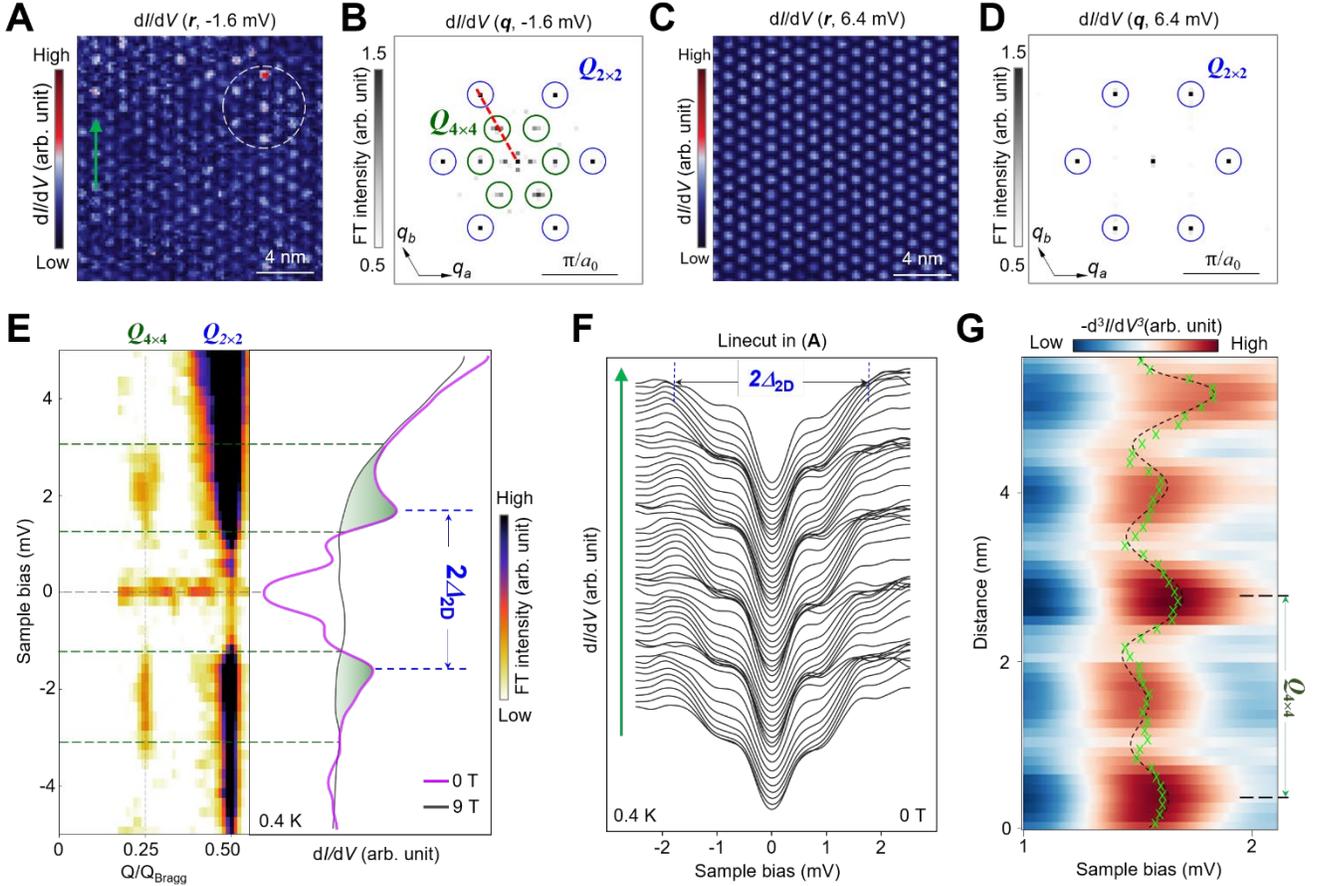

**Fig. 3. Observation of the 4×4 pair density wave modulation of $\Delta_{2D}$ at the 2×2 Cs ordered surface of $CsV_3Sb_5$.** (**A, B**) d$I$/d$V$ map at -1.6 mV (**A**) and the magnitude of the drift-corrected Fourier transform (**B**), showing the emergence of short-range ordered 4×4 spatial modulation (green circle in (**B**)). (**C, D**) d$I$/d$V$ map at 6.4 mV (**C**) and the magnitude of the drift-corrected Fourier transform (**D**) in the same region of (**A**), showing a uniform 2×2 lattice without 4×4 superlattice. (**E**), Fourier transform linecut of d$I$/d$V$ maps along one of three lattice direction (the red dash line in (**B**)) at 0.4 K as a function of sample bias, showing that ordering vectors at $Q_{4\times4}$ is nondispersive (left panel). The energy range of $Q_{4\times4}$ corresponds to the quasi-2D coherence as indicated by the comparison of d$I$/d$V$ spectra between 0 T and 9 T (right panel). Signals near $Q=0$ are manually set to 0 for clarity. (**F**) The waterfall plot of a d$I$/d$V$ linecut along green arrow in (**A**), showing spatial modulation of $\Delta_{2D}$. (**G**) The zoom-in intensity plot of -d$^3I$/d$V^3$ linecut corresponding to (**F**), highlighting that the spatial modulation of $\Delta_{2D}$ (green crosses) can be well described by two distinct periods of $2a_0$ and $4a_0$ (black dash curve), respectively. For all d$I$/d$V$ data in this figure, $V_{set}$= 50 mV, $I_t$=1.5 nA, $V_{mod}$=0.15 mV. The spectra in (**F**) are vertically shifted for clarity.

## Atomic condition and manipulation of quasi-2D superconductivity

There are important conditions for realizing the quasi-2D superconductivity. STM/S measurements throughout the sample surface show that not all the 2×2 Cs ordered regions show incipient band of bound states with the $P_{2D}$ close to the $E_F$ and pronounced SC gap $\Delta_{2D}$. The emergence of $\Delta_{2D}$ is sensitive to the size of 2×2 Cs ordered region. In small sized areas, $\Delta_{2D}$ is undetectable. In addition, even in the large-sized surface area, $\Delta_{2D}$ is occasionally much weaker than other areas (Fig. S14). To investigate the physical condition for the emergent quasi-2D superconductivity, we create *in situ* triangular shaped 2×2 Cs ordered islands with designed configurations by manipulating the Cs adatoms using the STM tip in the same surface region (Fig. 4A). Such engineered surface superstructures can avoid the possible spatial inhomogeneity of as-cleaved surface of $CsV_3Sb_5$ (Fig. 4A). More details are discussed in Methods and Fig. S15.

The first necessary condition for pronounced SC gap $\Delta_{2D}$ to appear is the critical size of 2×2 region must be larger than the coherence length of the quasi-2D superconductivity. Let's define the size of a region by the atom number $N$ on the edge of the isosceles triangle island with 2×2 Cs order (Fig. 4A). To minimize the edge induced quantum confinement effect (Fig. S16), we use the d$I$/d$V$ spectrum at the center of triangle to characterize the electronic states for each island. For the small-sized island ($N < 13$), we observe only the bulk SC gap $\Delta_{bulk}$ appearing below $T_{c, bulk}$ and small peaks around 5 mV arising from the pseudogap phase(*6, 42*) (Fig. 4B). When the island size is increased to $N = 13$, a pronounced peak $P_{2D}$ starts to be observable at 7.8 mV while $\Delta_{2D}$ is still undetectable. When the island size is further increased to $N > 19$, $\Delta_{2D}$ at around ±1.9 mV begins to be visible (red dash line in Fig. 4B). The critical size $N$ for emergent $\Delta_{2D}$ is about 20. Thus, as the nearest distance between two Cs atoms is about 1.1 nm, the critical edge length of triangle is about 20.9 nm. The observed size dependence of $P_{2D}$ and $\Delta_{2D}$ is summarized in Fig. 4C. Accordingly, the critical linear dimension of the quasi-2D superconductivity, defined as the distance from center to the edge of triangular island, is estimated to about 6 nm, which is indeed comparable to the SC coherence length estimated from the size of the vortex core.

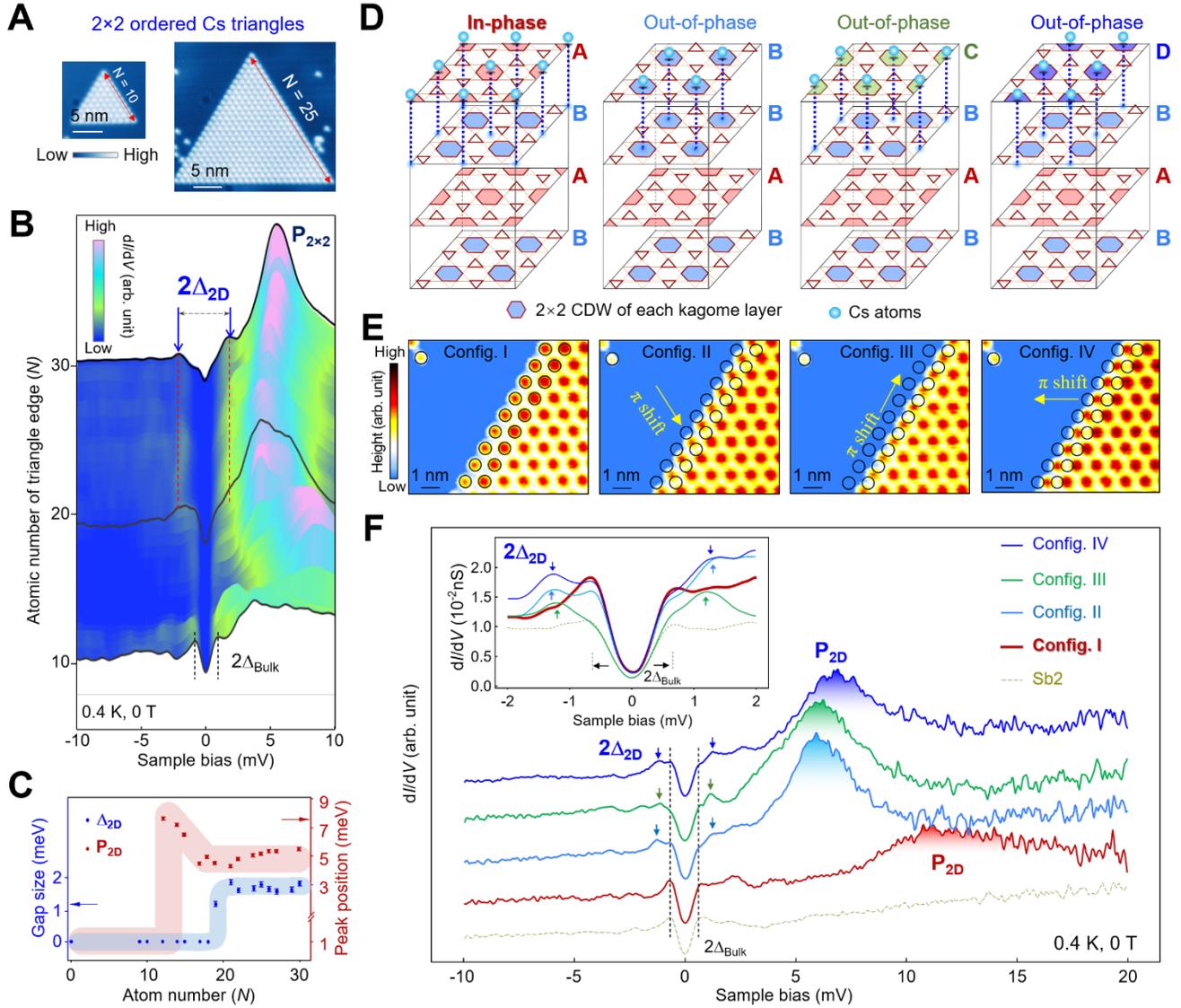

**Fig. 4. The correlation between $\Delta_{2D}$, size of 2×2 Cs ordered region and phase twist of bulk CDW in CsV$_3$Sb$_5$.** (**A**) Typical STM images showing two isosceles triangle islands with 2×2 Cs order. The size of the islands, defined by the atom number on the edge (red arrows), is $N=10$ and 25. $V_{set}$= 900 mV, $I_t$= 50 pA. (**B**) The d$I$/d$V$ spectra obtained at the center of triangle islands with increasing atom number $N$, showing the critical $N$ for the formation of $\Delta_{2D}$ (blue arrows) of 20. (**C**) The gap size of $\Delta_{2D}$ and energy position of $P_{2D}$ as a function of $N$, which shows an abrupt appearance of $\Delta_{2D}$ at $N = 20$ and $P_{2D}$ (~ 7.8 meV) at $N = 13$. The $P_{2D}$ gradually shift to a lower energy of ~ 5.5 meV at $N = 18$. (**D**) Schematic of four of 2×2 Cs ordered surface configurations by involving the correlation between the superlattice potential of the surface Cs order and 3D bulk CDW. One in-phase configuration fit with the out-of-plane stacking of the bulk CDW, namely ABAB, while the other three out-of-phase configurations interrupt the bulk CDW with a π-phase twist stacking fault at the surface, labelled as BBAB, CBAB and DBAB. (**E**) STM images showing four different 2×2 Cs terminations with the same size, constructed by atomic manipulation using the STM tip in the same surface region. $V_{set}$ = 900 mV, $I_t$ = 40 pA. (**F**) The d$I$/d$V$ spectra at the center of four configurations in (**E**), respectively, showing that the incipient in-gap density of states peak $P_{2D}$ in one of the four configurations locates at ~11.2 mV (red curve), much higher than other three (6.0-6.8 mV). Inset: zoom-in of d$I$/d$V$ spectra, showing that the feature of emergent $\Delta_{2D}$ for one configuration (red curve) is much weaker than others. For all d$I$/d$V$ data in this figure, $V_{set}$= 50 mV, $I_t$=1.5 nA, $V_{mod}$= 0.1 mV. The error bars in (**C**) represent the energy resolution in STS measurements.

The second condition involves the correlation between the superlattice potential of the surface Cs order and 3D bulk CDW in CsV$_3$Sb$_5$. As alluded above (Fig. 1A), each 2×2 unit cell in the surface region contains one Cs atom, which can occupy one of four equivalent sites located above the centers of the four hexagons in the 2×2 unit cell of the vanadium kagome lattice below. The four configurations are related by half a unit-cell translation or a π shift. This $Z_2 \times Z_2$ degeneracy is lifted when coupling to the bulk CDW is taken into account. In CsV$_3$Sb$_5$, the CDW has the periodicity of 2×2×2 at low temperatures as determined by STM(*18*) and scattering experiments(*21*). Although the precise nature is still under intensive investigation, it is believed to be a layer-stacking of bond ordered CDW with the 2×2 star-of-David (SD) or inverse-SD (ISD) configuration in the kagome plane(*28, 44*). We can therefore label the bulk CDW by a stacking sequence of ABABAB… along the c-axis, where A and B represent either two ISD (or two SD) configurations with a lateral π shift in one of the three lattice directions or a direct in-phase stacking of ISD and SD configurations. Since the ISD and the SD patterns are centered on one of the 4 hexagons in the 2×2 unit cell in the kagome plane, it is clear that only one of the four 2×2 Cs ordered configurations is in-phase with the out-of-plane stacking of the bulk CDW, while the other three interrupt the bulk CDW with a π-phase twist stacking fault at the surface. This three-to-one (3:1) ratio for the π-phase twisted to the in-phase terminations will be the same for all specific realizations of the 2×2×2 bulk CDW.

As an example, we consider the 2×2×2 CDW formed by stacking the ISD in the kagome plane with a lateral π shift in the out-of-plane direction(*17, 45*), which is the ground state in the density functional theory (DFT) calculations(*17*). We extend the DFT to explore how a 2×2 Cs ordered surface coupling to the bulk CDW (see Methods in Supplementary Materials). The results show that the 2×2 CDW in the top vanadium kagome plane always locks to the 2×2 Cs order on the surface, such that the center of the ISD pattern shifts to align directly underneath the occupied Cs site in the 2×2 unit cell (Fig. S2). This is due to the small binding energy (~10 meV) for the lateral π-phase shifted stacking of the ISD(*17*), compared to the large pinning energy due to the Cs atomic order. As a result, the possible $Z_2 \times Z_2$ Cs ordered configurations amount to four stacking terminations of the 2×2×2 CDW at the surface (Fig. 4D). Only one of them, labeled as ABAB, is in-phase with the bulk out-of-plane stacking of the ISD, while the other three interrupt the bulk CDW by a π-phase twist stacking fault at the topmost vanadium kagome layer and are labeled as BBAB, CBAB, and DBAB (Fig. 4D). We thus expect the emergence of different kinds of surface states with a 3:1 ratio that are localized in the Cs ordered regions.

To test this conjecture and gain microscopic insights, we experimentally construct the four different 2×2 Cs ordered terminations with the same size ($N = 25$ triangle) by atomic manipulation using the STM tip (Fig. 4E and Fig. S17). They are labelled as Configurations I-IV and are related by the marked half-unit cell π shift (Fig. 4E). Remarkably, the d$I$/d$V$ spectra in the center of each region show that the incipient in-gap density of states peak $P_{2D}$ at 6.0-6.8 mV and the SC gap $\Delta_{2D}$ emerge (cyan, green and blue curves) in Configurations II, III and IV, while $P_{2D}$ at about 11.2 mV and a much weaker feature of $\Delta_{2D}$ is observed in Config. I (red curve). The $P_{2D}$ indeed shows strong localization features around the 2×2 ordered Cs atoms in the d$I$/d$V$ linecuts (Fig. S3). Although STM is only sensitive to the top surface layer, the detected ratio of 3:1 supports that Configuration I corresponds to the in-phase ABAB termination while the other three correspond to π-phase twisted stacking fault of the bulk CDW at the surface (Fig. 4D). Therefore, the incipient narrow band $P_{2D}$ near E$_F$ is concurrent with $\Delta_{2D}$ (Fig. 1D), and the phase twisted stacking fault of the bulk CDW state at the surface, is the key condition for the emergent quasi-2D superconductivity that can be switched on and off by atomic manipulation of the Cs atoms.

**Discussions**

An important corollary of the mechanism is that in the absence of 3D bulk CDW order, 2×2 Cs surface reconstruction would be unable to generate the in-gap states on the surface and the quasi-2D superconductivity. To test this, we have grown single crystal of CsV$_{2.85}$Ti$_{0.15}$Sb$_5$, where long-range CDW order is suppressed(*46*) by the significant hole doping due to the Ti substitution of V. The spatially averaged d$I$/d$V$ spectra measured on differently reconstructed Cs surfaces, including 2×2 Cs ordered regions, as well as on the Sb terminated surface indeed show no significant differences; the incipient surface states $P_{2D}$ and the SC gap $\Delta_{2D}$ are both absent (Fig. S4). The absence of surface SC gap is further supported by the lack of intensity contrast in d$I$/d$V$ maps within (-2, 5) mV across the boundaries between 2×2 and other types of reconstructions (two exemplary maps are shown in Fig. S4). Conversely, we have studied single crystals of the other two members of the kagome $A$V$_3$Sb$_5$ family containing different alkali atoms, which also exhibit 3D bulk CDW order. We have indeed observed the incipient narrow band of states and quasi-2D SC gap on the 2×2 Rb and K ordered surface regions (Fig. S5). These observations attest to the robustness and consistency of our findings.

We reported the discovery of a novel quasi-2D superconductivity and a primary PDW on both naturally appearing and atomically engineered 2×2 Cs ordered surfaces of kagome metal CsV$_3$Sb$_5$. The transition temperature, SC energy gap, and the two-gap to $T_c$ ratio are significantly larger than those associated with the bulk SC state. The magnetic vortex exhibits an unprecedented spectrum of core states with a zero-energy plateau across the Fermi level. The $c$-axis upper critical field is an order of magnitude larger than that of the bulk SC state. Combining theoretical analysis and atomic manipulation of Cs order, we revealed the intriguing mechanism behind the unexpected correlation between the emergent 2D superconductivity and 3D bulk CDW order. The 2×2 Cs order behaves as an out-of-phase surface termination of the 2×2×2 bulk CDW, creating a narrow band of surface states inside the bulk CDW gap. The significantly enhanced density of state close to the Fermi level enabled to the first observation of a stronger superconductivity in the 2D limit of a kagome plane with the observed striking properties.

Although the as-cleaved 2×2 Cs surface reconstruction enabled our discovery and the atomic engineering of the Cs atoms allowed us to probe the mechanism, the underlying physics we revealed is much more general and its realizations are not limited to Cs surface reconstruction. For example, they can be created and manipulated using 2×2 superlattice potential and extended beyond the Cs surfaces. The π-phase dislocations can also happen in layered CDW state in the bulk at the stacking faults along the *c*-axis and influence the physical properties. Moreover, given the theoretical predictions for the bulk CDW state to be topological, either in the presence of time-reversal symmetry(*2, 17*)or when time-reversal symmetry is broken(*42*), it is conceivable that the surface states of the bulk CDW order and the emergent 2D superconductivity have intricate topological properties responsible for the observed unconventional vortex core spectrum.

Our findings significantly broaden the physical landscape of kagome metals and superconductors and push toward the 2D limit containing a single kagome lattice plane. They provide valuable insights for understanding the interplay between CDW order and superconductivity and a proof of principle for a novel pathway for atomic manipulation and topological defects engineering of quantum many-body states in kagome metals and other quantum materials.

# References

1. B. R. Ortiz *et al.*, New kagome prototype materials: discovery of $KV_3Sb_5$, $RbV_3Sb_5$, and $CsV_3Sb_5$. *Phys. Rev. Mater.* **3**, 094407 (2019).
2. B. R. Ortiz *et al.*, $CsV_3Sb_5$: A $Z_2$ topological kagome metal with a superconducting ground state. *Phys. Rev. Lett.* **125**, 247002 (2020).
3. S. Y. Yang *et al.*, Giant, unconventional anomalous Hall effect in the metallic frustrated magnet candidate, $KV_3Sb_5$. *Sci. Adv.* **6**, eabb6003 (2020).
4. Y. X. Jiang *et al.*, Unconventional chiral charge order in kagome superconductor $KV_3Sb_5$. *Nat. Mater.* **20**, 1353-1357 (2021).
5. H. Zhao *et al.*, Cascade of correlated electron states in the kagome superconductor $CsV_3Sb_5$. *Nature* **599**, 216-221 (2021).
6. H. Chen *et al.*, Roton pair density wave in a strong-coupling kagome superconductor. *Nature* **599**, 222-228 (2021).
7. L. P. Nie *et al.*, Charge-density-wave-driven electronic nematicity in a kagome superconductor. *Nature* **604**, 59-64 (2022).
8. C. Y. Guo *et al.*, Switchable chiral transport in charge-ordered kagome metal $CsV_3Sb_5$. *Nature* **611**, 461-466 (2022).
9. L. X. Zheng *et al.*, Emergent charge order in pressurized kagome superconductor $CsV_3Sb_5$. *Nature* **611**, 682-687 (2022).
10. Y. G. Zhong *et al.*, Nodeless electron pairing in $CsV_3Sb_5$-derived kagome superconductors. *Nature* **617**, 488-492 (2023).
11. T. Le *et al.*, Superconducting diode effect and interference patterns in kagome $CsV_3Sb_5$. *Nature* **630**, 64-69 (2024).
12. C. Mielke *et al.*, Time-reversal symmetry-breaking charge order in a kagome superconductor. *Nature* **602**, 245-250 (2022).
13. Q. W. Yin *et al.*, Superconductivity and normal-state properties of kagome metal $RbV_3Sb_5$ single crystals. *Chin. Phys. Lett.* **38**, 037403 (2021).
14. T.-H. Han *et al.*, Fractionalized excitations in the spin-liquid state of a kagome-lattice antiferromagnet. *Nature* **492**, 406-410 (2012).
15. J. X. Yin *et al.*, Giant and anisotropic many-body spin-orbit tunability in a strongly spin-orbit tunability in a strongly correlated kagome magnet. *Nature* **562**, 91-95 (2018).

16. D. F. Liu *et al.*, Magnetic Weyl semimetal phase in a kagomé crystal. *Science* **365**, 1282-1285 (2019).
17. H. X. Tan, Y. Liu, Z. Wang, B. Yan, Charge density waves and electronic properties of superconducting kagome metals. *Phys. Rev. Lett.* **127**, 046401 (2021).
18. Z. Liang *et al.*, Three-dimensional charge density wave and surface-dependent vortex-core states in a kagome superconductor $CsV_3Sb_5$. *Phys. Rev. X* **11**, 031026 (2021).
19. Y. Xing *et al.*, Optical manipulation of the charge-density-wave state in $RbV_3Sb_5$. *Nature*, (2024).
20. M. G. Kang *et al.*, Charge order landscape and competition with superconductivity in kagome metals. *Nat. Mater.* **22**, 186-193 (2023).
21. H. X. Li *et al.*, Observation of unconventional charge density wave without acoustic phonon anomaly in kagome superconductors $AV_3Sb_5$ (A = Rb, Cs). *Phys. Rev. X* **11**, 031050 (2021).
22. Y. Xiang *et al.*, Twofold symmetry of c-axis resistivity in topological kagome superconductor $CsV_3Sb_5$ with in-plane rotating magnetic field. *Nat. Commun.* **12**, (2021).
23. H. Li *et al.*, Rotation symmetry breaking in the normal state of a kagome superconductor $KV_3Sb_5$. *Nat. Phys.* **18**, 265-270 (2022).
24. H. Li *et al.*, Unidirectional coherent quasiparticles in the high-temperature rotational symmetry broken phase of $AV_3Sb_5$ kagome superconductors. *Nat. Phys.* **19**, 637-643 (2023).
25. B. Hu *et al.*, Robustness of the unidirectional stripe order in the kagome superconductor $CsV_3Sb_5$. *Chin. Phys. B* **31**, 058102 (2022).
26. P. Wu *et al.*, Unidirectional electron-phonon coupling in the nematic state of a kagome superconductor. *Nat. Phys.* **19**, 1143-1149 (2023).
27. T. Asaba *et al.*, Evidence for an odd-parity nematic phase above the charge-density-wave transition in a kagome metal. *Nat. Phys.* **20**, 40-46 (2024).
28. Y. S. Xu *et al.*, Three-state nematicity and magneto-optical Kerr effect in the charge density waves in kagome superconductors. *Nat. Phys.* **18**, 1470-1475 (2022).
29. D. R. Saykin *et al.*, High resolution polar kerr effect studies of $CsV_3Sb_5$: tests for time-reversal symmetry breaking below the charge-order transition. *Phys. Rev. Lett.* **131**, 016901 (2023).
30. C. Farhang, J. Y. Wang, B. R. Ortiz, S. D. Wilson, J. Xia, Unconventional specular optical rotation in the charge ordered state of Kagome metal $CsV_3Sb_5$. *Nat. Commun.* **14**, 5326 (2023).

31. J. Ge *et al.*, Charge-4e and charge-6e flux quantization and higher charge superconductivity in kagome superconductor ring devices. *Phys. Rev. X* **14**, 021025 (2024).

32. S. D. Wilson, B. R. Ortiz, AV$_3$Sb$_5$ kagome superconductors. *Nat. Rev. Mater.* **9**, 420-432 (2024).

33. C. Y. Guo *et al.*, Correlated order at the tipping point in the kagome metal CsV$_3$Sb$_5$. *Nat. Phys.* **20**, 579-584 (2024).

34. S.-W. Kim, H. Oh, E.-G. Moon, Y. Kim, Monolayer kagome metals AV$_3$Sb$_5$. *Nat. Commun.* **14**, 591 (2023).

35. B. Song *et al.*, Anomalous enhancement of charge density wave in kagome superconductor CsV$_3$Sb$_5$ approaching the 2D limit. *Nat. Commun.* **14**, 2492 (2023).

36. S. L. Ni *et al.*, Anisotropic superconducting properties of kagome metal CsV$_3$Sb$_5$. *Chin. Phys. Lett.* **38**, 057403 (2021).

37. T. Machida, Y. Kohsaka, T. Hanaguri, A scanning tunneling microscope for spectroscopic imaging below 90 mK in magnetic fields up to 17.5 T. *Rev. Sci. Instrum.* **89**, 093707 (2018).

38. Z. H. Huang *et al.*, Tunable vortex bound states in multiband CsV$_3$Sb$_5$-derived kagome superconductors. *Sci. Bull.* **69**, 885-892 (2024).

39. Z. Q. Gao, Yu-Ping Lin, D.-H. Lee, Pair-breaking scattering interference as a mechanism for superconducting gap modulation. arXiv: 2310.06024 [Preprint] (2023).

40. T. Wei, Y. Liu, W. Ren, Z. Wang, J. Wang, Observation of cooper-pair density modulation state. arXiv: 2404.16683 [Preprint] (2024).

41. L. Kong *et al.*, Observation of intra-unit-cell superconductivity modulation. arXiv: 2404.10046 [Preprint] (2024).

42. S. Zhou, Z. Wang, Chern Fermi pocket, topological pair density wave, and charge-4e and charge-6e superconductivity in kagomé superconductors. *Nat. Commun.* **13**, 7288 (2022).

43. B. Hu *et al.*, Robustness of the unidirectional stripe order in the kagome superconductor CsV$_3$Sb$_5$. *Chin. Phys. B* **31**, 058102 (2022).

44. B. R. Ortiz *et al.*, Fermi surface mapping and the nature of charge-density-wave order in the kagome superconductor CsV$_3$Sb$_5$. *Phys. Rev. X* **11**, 041030 (2021).

45. M. H. Christensen, T. Birol, B. M. Andersen, R. M. Fernandes, Theory of the charge density wave in AV$_3$Sb$_5$ kagome metals. *Phys. Rev. B* **104**, 214513 (2021).

46. H. T. Yang *et al.*, Titanium doped kagome superconductor CsV$_{3-x}$Ti$_x$Sb$_5$ and two distinct phases. *Sci. Bull.* **67**, 2176-2185 (2022).